\documentclass[aps,prd,10pt,nofootinbib,twocolumn,eqsecnum,showpacs,showkeys,superscriptaddress,preprintnumbers,altaffilletter]{revtex4-2}
\usepackage{graphicx}
\usepackage{hyperref}
\usepackage{subcaption}
\usepackage{amsmath}
\usepackage{multirow}
\usepackage{makecell}
\usepackage{dcolumn}
\usepackage{amssymb}
\usepackage{amsfonts}
\usepackage{amsbsy}

\begin{document}

\title{Searching for Extra Dimensions with Gravitational Waves: Dark-Siren Constraints from GWTC-4}

\date{\today}

\author{Anson Chen}
\email{chena@ucas.ac.cn}  
% \altaffiliation{chena@ucas.ac.cn}
\affiliation{International Center for Theoretical Physics Asia-Pacific, University of Chinese Academy of Sciences, 100190 Beijing, China}

\author{Jun Zhang}
%\altaffiliation{Kitt Peak National Observatory}
\affiliation{International Center for Theoretical Physics Asia-Pacific, University of Chinese Academy of Sciences, 100190 Beijing, China}
\affiliation{Taiji Laboratory for Gravitational Wave Universe, University of Chinese Academy of Sciences, 100049 Beijing, China}

\begin{abstract}
Higher-dimensional theories of gravity predict that gravitational waves (GWs) can propagate into extra spatial dimensions, leading to modified amplitude damping over cosmological distances. Measurements of GW luminosity distances therefore provide a unique probe of the dimensionality of spacetime. In this work, we constrain higher-dimensional GW propagation using the dark-siren method with the Gravitational-Wave Transient Catalog 4.0 (GWTC-4). We adopt a phenomenological parameterization motivated by braneworld scenarios, in which deviations from General Relativity are characterized by the spacetime dimension number $D$ and a crossover scale $R_c$ governing the transition between four- and higher-dimensional gravity. We perform a hierarchical Bayesian analysis combining 141 compact binary coalescences from GWTC-4 with line-of-sight galaxy information from the GLADE+ catalog. For a prior $H_0 \in [65,77]\ {\rm km~s^{-1}Mpc^{-1}}$ and $\log(R_c/{\rm Mpc}) \in [2.7,4.0]$, we obtain $D = 4.38^{+1.91}_{-1.01}$ (68\% credible interval). We also find that the inferred posterior distribution of $R_c$ accumulates near the upper prior boundary, indicating that the crossover scale remains poorly constrained by current observations. We further show that the inferred constraint on $D$ depends sensitively on the assumed prior range of $R_c$, which determines the characteristic distance scale at which deviations from General Relativity become significant. Our results provide the first GWTC-4 dark-siren constraints on higher-dimensional GW propagation and demonstrate that current observations remain consistent with four-dimensional General Relativity.
\end{abstract}

\maketitle

\section{Introduction}

The quest to understand gravity beyond the framework of four-dimensional spacetime has led to the development of higher-dimensional theories, which posit that the universe may contain additional spatial dimensions beyond those directly observable. These ideas trace back to early unification attempts by Kaluza and Klein, who demonstrated that extending spacetime could naturally incorporate electromagnetism within a higher-dimensional gravitational framework \cite{Kaluza:1921tu,Klein:1926tv}. In modern contexts, higher-dimensional gravity arises prominently in approaches such as string theory and braneworld scenarios, where extra dimensions are not merely mathematical artifacts but essential components required for internal consistency \cite{Green:1987sp,Witten:1995ex,Polchinski:1998rq}.

A central motivation for considering higher-dimensional models is their potential to address long-standing challenges in theoretical physics, including the hierarchy problem, the nature of dark matter and dark energy, and the unification of gravity with quantum mechanics. In particular, models with large or warped extra dimensions have shown how the fundamental Planck scale could be lowered to the electroweak scale, providing a novel perspective on the hierarchy problem \cite{Arkani-Hamed:1998jmv,Randall:1999ee}. By allowing gravitational degrees of freedom to propagate in additional dimensions, these frameworks often predict modifications to gravitational behavior at both short and cosmological scales. Such deviations open avenues for experimental and observational tests, ranging from collider experiments to precision measurements of Newtonian gravity and astrophysical observations \cite{Antoniadis:1998ig,Csaki:2004ay,Adelberger:2003zx}.

% Despite their conceptual appeal, higher-dimensional theories face significant challenges, particularly in explaining why extra dimensions are not directly observed. Mechanisms such as compactification and localization have been proposed to reconcile this discrepancy, suggesting that additional dimensions may be either compact and extremely small or accessible only to specific fields \cite{Klein:1926tv,Rubakov:1983bb}. In string-theoretic contexts, the geometry and topology of the compact extra dimensions play a crucial role in determining the low-energy effective physics, leading to a rich landscape of possible solutions \cite{Green:1987sp,Giddings:2001yu}.

An increasingly powerful avenue for testing higher-dimensional theories of gravity arises from the observation of gravitational waves. In models with extra dimensions, gravitational waves (GWs) may propagate partly into the higher-dimensional bulk, leading to modifications in their amplitude, dispersion, and polarization compared to predictions from four-dimensional general relativity. For instance, energy leakage into extra dimensions can result in an anomalous damping of the waveform, while the presence of additional graviton modes may induce deviations in the propagation speed or generate characteristic spectral distortions. Observations from ground-based interferometers and future space-based detectors therefore provide a sensitive probe of such effects, allowing constraints to be placed on the size, geometry, and dynamics of extra dimensions \cite{Deffayet:2007kf,Nishizawa:2017nef,Visinelli:2017bny,Kwon:2019gsa,Corman:2020pyr,Corman:2021avn}.

In this paper, we focus on tests of higher-dimensional theories of gravity via GW propagation effects using the dark siren approach. Proposed by Schutz in 1986 \cite{Schutz1986}, the dark siren method constrains cosmological parameters by associating GW luminosity distances with galaxy redshifts within a Bayesian statistical framework. In the past few years, mature toolkits for dark siren analysis have been well-developed \cite{Gray:2019ksv,Gray:2021sew,Gray:2023wgj,Mastrogiovanni:2023emh,Mastrogiovanni:2023zbw,Borghi:2023opd,Tagliazucchi:2025ofb,Jin:2025dvf,Song:2025bio,Song:2025ddm}. Using GWTC-4 data from LIGO-Virgo-KAGRA (LVK), the constraint on Hubble constant $H_0$ by combining the bright siren GW170817 and dark siren analysis with 141 compact binary coalescences (CBCs) yields $H_0 = 76.6^{+13.0}_{-9.5} {\rm km~s^{-1}~Mpc^{-1}}$ \cite{LIGOScientific:2025jau} in the Lambda Cold Dark Matter (LCDM) cosmology model. In addition, dark siren analysis also constrains modified gravity parameterizations via GW propagation effects \cite{Mukherjee:2020mha,Finke:2021aom,Mancarella:2021ecn,Ezquiaga:2021ayr,Leyde:2022orh,Chen:2023wpj,Romano:2022jeh,Romano:2023ozy,Romano:2024apw,Afroz:2024joi,Zhang:2024rel,Chen:2024xkv,Tagliazucchi:2025ofb,LIGOScientific:2025jau}. Most measurements constrain the general parameterization using $(\Xi_0,n)$ parameters, and Horndeski-class parameterization with the assumption of $\alpha_M(z)=c_M\Omega_\Lambda(z)/\Omega_{\Lambda,0}$ within the LCDM background. But several studies have also constrained the higher-dimensional theories with GW170817 \cite{Pardo:2018ipy,LIGOScientific:2018dkp}, and with binary black hole (BBH) mergers in GWTC-3 \cite{MaganaHernandez:2021zyc,Leyde:2022orh,Chen:2023wpj}, finding that the spacetime dimension remains consistent with 4, as predicted by General Relativity (GR). %In this work, we aim to extend the test of higher-dimensional theories with full-population dark siren analysis using the latest GWTC-4 data \cite{LIGOScientific:2025slb}, associated with the GLADE+ galaxy catalog \cite{Dalya:2021ewn}. Moreover, we employ the \texttt{gwcosmo} package accelerated with tensor-based computations using graphics processing units (GPUs)\cite{Papadopoulos:2026puy}, which greatly reduces the computational time required for Bayesian inference over 141 CBCs.

In this work, we aim to extend the test of higher-dimensional theories with full-population dark siren analysis by the \texttt{gwcosmo} package using the latest GWTC-4 data \cite{LIGOScientific:2025slb}, associated with the GLADE+ galaxy catalog \cite{Dalya:2021ewn}. This work differs from the previous GWTC-3 analysis \cite{Chen:2023wpj} in two main respects. First, the GWTC-4 dataset used here includes events detected during the LVK O4a observing run in addition to those in GWTC-3. We also adopt a different event-selection threshold, requiring a false alarm rate (FAR) of $<0.25~\mathrm{yr}^{-1}$. The resulting sample comprises 137 binary black hole (BBH) mergers and four events involving neutron stars. By adopting a full-population mass model spanning both neutron-star and black-hole masses, we analyze all 141 CBCs within a single Bayesian framework, rather than treating BBH and neutron-star–black-hole mergers separately as in the GWTC-3 analysis \cite{Chen:2023wpj}. Second, we use the new version of the \texttt{gwcosmo} package accelerated through tensor-based computations on graphics processing units (GPUs) \cite{Papadopoulos:2026puy}. This substantially reduces the computational time required for Bayesian inference with all 141 CBCs while yielding results consistent with the original implementation. The GLADE+ galaxy catalog, however, remains unchanged from that used in the GWTC-3 analysis \cite{Chen:2023wpj}.

The structure of this work are given as the following. The GW propagation effect induced by higher-dimensional theories is first reviewed in Section \ref{sec:theory}, and then the hierarchical Bayesian framework of dark siren analysis is introduced in Section \ref{sec:framework}. Next, the data and the inference setup we use are presented in Section \ref{sec:data}. Finally, the results are given in Section \ref{sec:results}, followed by the conclusions in Section \ref{sec:conclusions}.

\section{GW propagation effect in higher-dimensional theories}
\label{sec:theory}

In higher-dimensional theories, such as the Dvali-Gabadadze-Porrati (DGP) theory \cite{Dvali:2000hr}, gravitational interactions extend into higher dimensions beyond the 4-dimensional spacetime described in GR, while ordinary matter is confined to a 3-dimensional spatial brane. This configuration results in an effective ``leakage" of gravity over cosmological distance scales, which translates into modified GW propagation featuring an additional damping effect. Here we briefly review the derivation of the resulting modification in GW luminosity distance following Ref. \cite{Corman:2021avn}. 

Considering GWs as linearized tensor perturbations propagating in flat spacetime, in the harmonic gauge we have \cite{2007gwte.book.....M}
\begin{equation}
    \Box h_{\mu\nu}=0.
\end{equation}
The GW waveform can be described by
\begin{equation}
    h_{\mu\nu}=e_{\mu\nu}Ae^{i\Phi/\epsilon},
\end{equation}
where $A$ and $\Phi$ are the amplitude and the phase of GW, $e_{\mu\nu}$ is the polarization, and $\epsilon$ represents a small quantity measuring the short wavelength so that the phase can be expanded over $\epsilon$ in the geometric optics limit. The leading orders of the expansion give
\begin{equation}
    k_\mu k^\mu = 0,
\end{equation}
\begin{equation}
    \nabla_\mu (A^2 k^\mu) = 0,
    \label{eq:conservation}
\end{equation}
where $k^\mu$ is the wave vector. The first equation describes GW propagating along null geodesics, and the second equation represents the conservation of GW flux energy.

Now considering a Minkowski spacetime with $D=N+1$ dimension, the metric in hyperspherical coordinates can be given by
\begin{equation}
    ds^2=-dt^2+dr_N^2+r_N^2 d\Omega_{N-1}^2,
\end{equation}
where $r_N$ is the radius in hyperspherical coordinates. For a radially propagating wave with the speed of light ($c=1$),
we have $k^\mu=(\omega,\omega,0,\dots,0)$. Making use of the transformation 
\begin{equation}
    \nabla_\mu (A^2 k^\mu) = \frac{1}{\sqrt{-g}}\partial_\mu\left(\sqrt{-g}\,A^2 k^\mu\right),
\end{equation}
and in hyperspherical coordinates $\sqrt{-g}=r_N^{D-2}$, equation (\ref{eq:conservation}) leads to
\begin{equation}
    \partial_t(A^2\omega) + \frac{1}{r_N^{D-2}}\partial_{r_N}\left(r_N^{D-2}A^2\omega\right)=0.
\end{equation}
In the geometric optics limit, the amplitude changes slowly compared with the oscillation period, so the time derivative term can be neglected. This leaves
\begin{equation}
    \partial_{r_N}\left(r_N^{D-2}A^2\right)=0.
\end{equation}
Integrating this with respect to $r_N$ gives
\begin{equation}
    A\propto r_N^{-(D-2)/2}.
\end{equation}
It shows that the distance of GW is scaled differently in higher-dimensional theories. Assuming that this new distance scaling is the only modification to GW waveform, the CBC waveform amplitude in the source frame yields
\begin{equation}
    A\propto \frac{{\cal M}_c^{5/3}}{r_N^{(D-2)/2}} f_s^{2/3},
\end{equation}
where ${\cal M}_c$ is the chirp mass. Generalizing this to the Friedmann--Lemaître--Robertson--Walker (FLRW) metric, the metric of the hyperspherical radius is scaled by $a(t)$, so following the same derivation we obtain
\begin{equation}
    A\propto (a(t)r_N)^{-(D-2)/2}.
\end{equation}
In the detector frame, $a(t)=a_0$. Rewriting chirp mass and observed frequency with redshifted quantities, we have
\begin{equation}
    A\propto \frac{{\cal M}_{cz}^{5/3}}{(1+z)(a_0 r_N)^{(D-2)/2}} f_o^{2/3},
\end{equation}
from which we can define the GW luminosity distance as
\begin{equation}
    d^{\rm GW}_L \propto (1+z)(a_0 r_N)^{(D-2)/2}
\end{equation}
% Given the EM luminosity distance in higher-dimensional theories \cite{1992ApJ...397....1C}
% \begin{equation}
%     d_L^{(D)} = a_0 r_N (1+z)^{2/(D-2)}
% \end{equation}

On the other hand, since photons are assumed to propagate only in four dimensions, the EM luminosity distance is
\begin{equation}
    d_L^{\mathrm{EM}}=a_0 r_3(1+z),
\end{equation}
The crucial link between EM and GW signals is that they both follow null geodesics and therefore travel the same comoving radial distance $r_N=r_3$ \cite{Corman:2021avn}. Using this equality, the higher-dimensional GW luminosity distance becomes
% \begin{equation}
%     d_L^{(D)} = d_L^{\mathrm{EM}}(1+z)^{\frac{4-D}{D-2}}.
% \end{equation}
% Substituting this into the GW amplitude scaling gives
\begin{align}
    d^{\rm GW}_L & \propto (1+z)(a_0 r_3)^{(D-2)/2} \nonumber\\
    & \propto d_L^{\mathrm{EM}}\left(\frac{d_L^{\mathrm{EM}}}{1+z}\right)^{(D-4)/2}.    
\end{align}
This implies that for $D=4$, the luminosity distance reduces to $d^{\rm GW}_L=d^{\rm EM}_L$. 

\begin{figure}
    \centering
    \includegraphics[width=\linewidth]{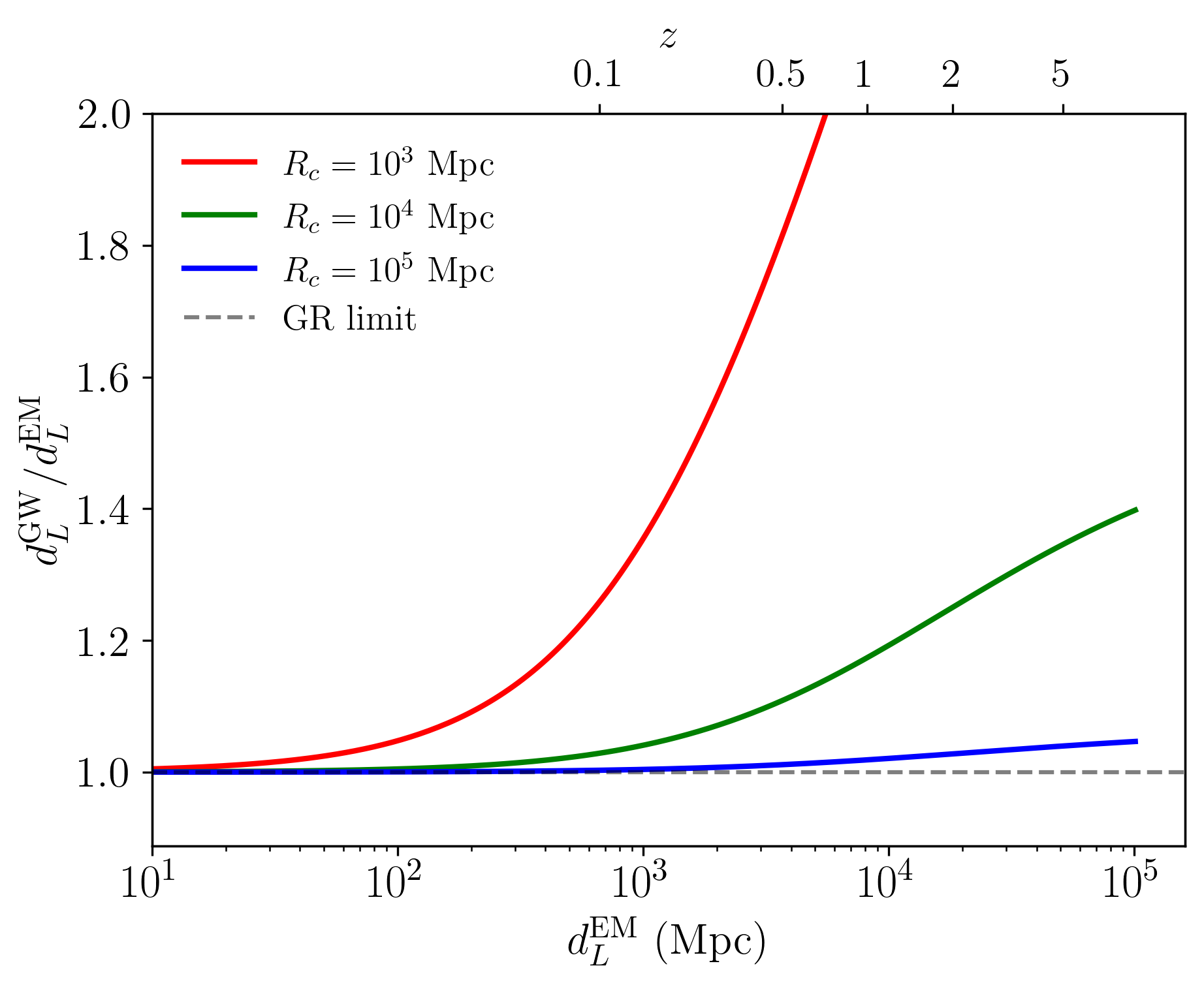}
    \caption{The ratio of $d^{\rm GW}_L/d^{\rm EM}_L$ plotted against $d^{\rm EM}_L$ for $D=5$ and $n=1$ with different values of $R_c$.}
    \label{fig:dgp_ratio_plot}
\end{figure}

In the framework of the DGP theory, graviton is a resonance of width $1/R_c$, with $R_c$ representing the crossover scale (See definition of $R_c$ in Appendix \ref{app:DGP}). On distances significantly shorter than $R_c$, resonance effects are suppressed, and the dynamics are effectively described by 4-dimensional GR on the brane. Conversely, at distances larger than $R_c$, gravity must be described in the full 5-dimensional bulk. This motivates a phenomenological parameterization for $d^{\rm GW}_L$ considering the crossover effect \cite{Deffayet:2007kf,Corman:2020pyr}, which eventually yields
\begin{equation}
    d_L^{\mathrm{GW}} = d_L^{\mathrm{EM}} \left[1+\left(\frac{d_L^{\mathrm{EM}}}{R_c(1+z)}\right)^n\right]^{\frac{D-4}{2n}}.
    \label{eq:dL_GW}
\end{equation}
Here $n$ is a phenomenological parameter. For $d_L^{\mathrm{EM}} \ll R_c(1+z)$, it reduces to $d^{\rm GW}_L=d^{\rm EM}_L$. While for $d_L^{\mathrm{EM}} \gtrsim R_c(1+z)$, the deviation between $d^{\rm GW}_L$ and $d^{\rm EM}_L$ becomes significant for $D\neq 4$. For $d^{\rm EM}_L$ defined in the usual 4-dimensional spacetime, we adopt the $\Lambda$CDM cosmology background expression:
\begin{equation}
    d^{\rm EM}_L(z) = \frac{c(1+z)}{H_0}\int_0^z \frac{{\rm d}z'}{ \sqrt{\Omega_{m,0}(1+z)^3 + \Omega_{\Lambda,0}}}.
    \label{eq:dL_EM}
\end{equation}
Combining equation \eqref{eq:dL_GW} and \eqref{eq:dL_EM}, this new $d^{\rm GW}_L$--$z$ relation is the key to test higher-dimensional theories through dark siren analysis. 
A demonstration figure of the ratio $d^{\rm GW}_L/d^{\rm EM}_L$ aginst $d^{\rm EM}_L$ for $D=5$ in the DGP theory is shown in Fig. \ref{fig:dgp_ratio_plot}. The deviation of the ratio from GR takes place at different distance scales for different values of $R_c$. Therefore, the sensitivity of the distance ratio on the spacetime dimension number $D$ depends heavily on $R_c$.

\section{Hierarchical Bayesian framework for dark siren analysis}
\label{sec:framework}

In this section, we briefly describe the hierarchical Bayesian framework for dark siren analysis used in our work. Since this framework has been introduced explicitly in many previous works \cite{Mandel:2018mve,Vitale:2020aaz,Ezquiaga:2022zkx,Mastrogiovanni:2023emh,Mastrogiovanni:2023zbw,Gray:2023wgj,LIGOScientific:2025jau}, we refer readers to the cited papers for more details. Specifically, we will use the \texttt{gwcosmo} implementation of the framework \cite{Gray:2019ksv,Gray:2021sew,Gray:2023wgj} for our analysis.

Our goal is to obtain the posterior probability of cosmological hyperparameters $\Lambda_c$ from the Bayesian framework. In our analysis, $\Lambda_c$ includes the background parameters like $H_0$, and the phenomenological parameters in $d^{\rm GW}_L$ for higher-dimensional theories. In specific, we consider free parameters to be $\Lambda_c=\{H_0,D,\log R_c\}$. We fix $\Omega_{m,0}$ as it is difficult to measure with current GW data, and we adopt a more precise measurement value $\Omega_{m,0}=0.3065$ from the Planck mission \cite{Planck:2015fie}. We also fix $n=1$ as it is weakly constrained from dark siren analysis \cite{Chen:2023wpj}. Apart from $\Lambda_c$, we jointly constrain hyperparameters $\Lambda$ in GW population model in hierarchical analysis, over which we will marginalize to obtain constraint on $\Lambda_c$. For the ensemble of GW data $\{d\}$ from $N_\text{det}$ detected events, the posterior probability of hyperparameters $\{\Lambda_c,\Lambda\}$ is given by \cite{Mandel:2018mve,Vitale:2020aaz}
\begin{equation}
\label{eq:posterior}
\begin{split}
    p\left( \Lambda_c, \Lambda | \{d\}, N_{\mathrm{det}} \right) & \propto  \pi(\Lambda_c)\, \pi(\Lambda) \,p(N_{\mathrm{det}} | \Lambda_c,\Lambda) \\
    & \times  \xi(\Lambda_c, \Lambda )^{-N_{\mathrm{det}}} \, p(\{d\}|\Lambda_c,\Lambda)
\end{split}
\end{equation}
Here $\pi$ represents the priors of hyperparameters, and $p(N_{\mathrm{det}} | \Lambda_c,\Lambda)$ denotes the probability to detect $N_{\mathrm{det}}$ events. We adopt the assumption that $p(N_{\mathrm{det}} | \Lambda_c)$ is invariant of hyperparameters when choosing a prior of $1/R$ when marginalizing over the intrinsic merger rate $R$ \cite{Fishbach:2018edt,Mandel:2018mve}. The likelihood $p(\{d\}|\Lambda_c,\Lambda)$ is obtained by marginalizing over the detector-frame waveform parameters $\theta^\mathrm{det}_{i}$ from parameter estimation for the $i$-th event,
\begin{equation}
\begin{split}
    &p(\{d\}|\Lambda_c,\Lambda) = \prod_{i=1}^{N_{\mathrm{det}}} \int \! \! \mathrm{d}\theta^\mathrm{det}_{i}\frac{p\big( {\theta}^\mathrm{det}_{i} | d_i \big)}{\pi_{\rm PE}(\theta^\mathrm{det}_{i})} \, \\
    & \times \left[ \Big|\frac{\mathrm{d}\theta^\mathrm{det}_{i}(\theta,\Lambda_c)}{\mathrm{d} \theta_i} \Big|^{-1} p_{\mathrm{pop}}(\theta_i | \Lambda_c,\Lambda)\right]_{\theta_i=\theta_{ i}(\theta^\mathrm{det}_{i}, \Lambda_c)} \, .
    \label{eq:marginal_lik}
\end{split}
\end{equation}
In cosmological analyses, the waveform parameter set is usually taken as $\theta^\mathrm{det}_{i}=\{m^\mathrm{det}_{1,i},m^\mathrm{det}_{2,i},d^{\rm GW}_{L,i}\}$, where $m^\mathrm{det}_{1}$ and $m^\mathrm{det}_{2}$ denote the primary and the secondary mass of the binary. The prior of posterior samples from parameter estimation is uniform on the component mass, $\pi_{\rm PE}(m^\mathrm{det}_{1,2})\propto 1$, while the prior on $d^{\rm GW}_L$ is $\pi_{\rm PE}(d^{\rm GW}_L)\propto (d^{\rm GW}_L)^2$. In addition, $p_{\mathrm{pop}}(\theta_i | \Lambda)$ represents the GW population prior modelled with source-frame parameters $\theta_{i}=\{m_{1,i},m_{2,i},z_{i}\}$, where $m_{1,2}=m^\mathrm{det}_{1,2}/[1+z(d^{\rm GW}_L,\Lambda_c)]$ Therefore, the Jacobian transformation from the source frame to the detector frame is given by
\begin{equation}
    \Big|\frac{\mathrm{d}\theta^\mathrm{det}_{i}(\theta,\Lambda_c)}{\mathrm{d} \theta_i} \Big| = (1+z)^2\frac{\mathrm{d}d^{\rm GW}_L(z,\Lambda_c)}{\mathrm{d}z}.
\end{equation}
Meanwhile, $\xi(\Lambda_c, \Lambda )$ represents the selection effects given the detection threshold, which is expressed as
\begin{equation}
\begin{split}
    &\xi(\Lambda_c, \Lambda )= \int \! \mathrm{d} \theta^\mathrm{det} \,  P({\rm det} | {\theta}^\mathrm{det}) \, \\
& \times \left[ \Big|\frac{\mathrm{d}\theta^\mathrm{det} (\theta,\Lambda_c)}{\mathrm{d \theta}} \Big|^{-1} p_{\mathrm{pop}}(\theta | \Lambda_c,\Lambda)\right]_{\theta=\theta(\theta^\mathrm{det}, \Lambda_c)} ,
\end{split}
\end{equation}
where $P({\rm det} | {\theta}^\mathrm{det}) \in [0,1]$ denotes whether an event with ${\theta}^\mathrm{det}$ is detected or not. The selection effects are generated with a large number of event injections in the parameter space of ${\theta}^\mathrm{det}$ given a selection criterion. In this work, we adopt the same 141 CBCs in GWTC-4 with a false alarm rate (FAR) $<0.25~\mathrm{yr}^{-1}$ as in GWTC-4 cosmological analysis. Therefore, we use the same set of injections as in GWTC-4 cosmological analysis as well (see Appendix A of Ref. \cite{LIGOScientific:2025jau} for injection details). 

The GW source population prior can be divided into mass distribution and redshift distribution. The redshift distribution is further contributed by the intrinsic astrophysical merger rate $\psi(z|\Lambda)$ in unit space and time, and the host galaxy redshift distribution along the line-of-sight (LOS). In the \texttt{gwcosmo} framework, the population prior is computed pixel by pixel as
\begin{equation}
    p_{\mathrm{pop}}(\theta | \Omega_j, \Lambda_c,\Lambda) = p(m_1,m_2|\Lambda)\frac{\psi(z|\Lambda)}{(1+z)} p(z|\Omega_j,\Lambda_c),
\end{equation}
where $\Omega_j$ denotes the direction of the $j$-th pixel within the GW localization area. The likelihood in equation \eqref{eq:marginal_lik} is then computed by summing over all $N_{\mathrm{pix}}$ pixels within the GW localization area,
\begin{equation}
\begin{split}
    &p(\{d\}|\Lambda_c,\Lambda) = \prod_{i=1}^{N_{\mathrm{det}}} \int \! \! \mathrm{d}\theta^\mathrm{det}_{i} \sum_j^{N_{\mathrm{pix}}} \frac{p\big( {\theta}^\mathrm{det}_{i} | \Omega_j,d_i \big)}{\pi_{\rm PE}(\theta^\mathrm{det}_{i})} \, \\
    & \times \left[ \Big|\frac{\mathrm{d}\theta^\mathrm{det}_{i}(\theta,\Lambda_c)}{\mathrm{d} \theta_i} \Big|^{-1} p_{\mathrm{pop}}(\theta_i | \Omega_j,\Lambda_c,\Lambda)\right]_{\theta_i=\theta_{ i}(\theta^\mathrm{det}_{i}, \Lambda_c)} \, .
    \label{eq:marginal_lik_pixel}
\end{split}
\end{equation}
In addition, the selection effect is pixelized as well. However, under the assumption that GW sources distribute isotropically across the sky, $P({\rm det} | {\theta}^\mathrm{det})$ remains invariant over all pixels, so we have
\begin{equation}
\begin{split}
    \xi(\Lambda_c, \Lambda ) &= \int \! \mathrm{d} \theta^\mathrm{det} \,  P({\rm det} | {\theta}^\mathrm{det}) \bigg[ \Big|\frac{\mathrm{d}\theta^\mathrm{det} (\theta,\Lambda_c)}{\mathrm{d \theta}} \Big|^{-1} \, \\
    & \times \sum_j^{N_{\mathrm{pix}}}p_{\mathrm{pop}}(\theta | \Omega_j, \Lambda_c,\Lambda)\bigg]_{\theta=\theta(\theta^\mathrm{det}, \Lambda_c)}.
\end{split}
\end{equation}
For each pixel, the LOS redshift prior $p(z|\Omega_j,\Lambda_c)$ consists of two parts. The first part is the in-catalog part constructed with a galaxy catalog. Each galaxy along the LOS is represented by a narrow Gaussian peak in redshift distribution, weighted by its luminosity $L$. Given that brighter galaxies have higher mass, they are more likely to host binary mergers. The weight usually takes the form of $w=(L/L_*)^\epsilon$, where $\epsilon=1$ and $\epsilon=0$ represent the luminosity-weighted case and the uniform-weighted case respectively.
But since the galaxy catalog is incomplete over higher redshift, the LOS prior is compensated by the out-of-catalog part, which assumes that galaxies below the observed magnitude threshold follow the uniform-in-comoving-volume distribution. Interested readers are refereed to Ref. \cite{Gray:2023wgj} for more details in LOS prior construction.

\section{Inference setup}
\label{sec:data}

Following the GWTC-4 cosmology paper \cite{LIGOScientific:2025jau}, we run the dark siren analysis with 141 CBCs in GWTC-4 \cite{LIGOScientific:2025slb} with FAR$<0.25~\mathrm{yr}^{-1}$, including 137 BBHs and 4 binary mergers in which at least one component is a neutron star. We adopt the same posterior samples from parameter estimation (PE) as well, which are generated using the \textsc{IMRPhenomXPHM} waveform model \cite{Pratten:2020ceb} for events from O1-O3 runs and the \textsc{IMRPhenomXPHM\_SpinTaylor} waveform model \cite{Colleoni:2024knd} for O4a events of the $9^{\rm th}$ Stable Release. We also exclude the event GW231123\_135430 as it suffers from PE systematics in waveform choices.

We use the \textsc{FullPop}-4.0 mass model and the \textsc{MultiPeak} mass model from the GWTC-4 population analysis paper \cite{LIGOScientific:2025pvj,Fishbach:2020ryj,Farah:2021qom,Mali:2024wpq} to build the mass prior $p(m_1,m_2|\Lambda)$. These mass models feature two Gaussian peaks on top of a power-law distribution for the primary component mass in binary systems. While the \textsc{MultiPeak} model only describes black hole mass distribution with a minimum mass cutoff, the \textsc{FullPop}-4.0 model extend its lower mass bound to $\sim1M_\odot$ to describe the total mass distribution of neutron stars and black holes with a gap between them. In addition, we adopt the merger rate redshift evolution model $\psi(z|\Lambda)$ inspired by the Madau-Dickinson star formation rate, which is constructed by two power-law distributions with a turn-over point around $z \simeq 2$.
The hyperparameters in these population models and their priors used in our analysis are listed in Appendix \ref{app:pop_mod}.

We use the GLADE+ galaxy catalog \cite{Dalya:2021ewn} in our dark siren analysis, which is also used in the GWTC-4 paper. GLADE+ is an all-sky galaxy catalog containing approximately 22 million galaxies, constructed by combining six astronomical datasets: the Gravitational Wave Galaxy Catalog (GWGC) \cite{White:2011qf}, HyperLEDA \cite{Makarov:2014txa}, the 2 Micron All-Sky Survey Extended Source Catalog (2MASS XSC, \cite{2MASS:2006qir}), the 2MASS Photometric Redshift Catalog (2MPZ) \cite{Bilicki:2013sza}, the WISExSCOS Photometric Redshift Catalog (WISExSCOSPZ) \cite{Bilicki:2016irk} and the Sloan Digital Sky Survey quasar catalog from the 16th data release (SDSS-DR16Q) \cite{eBOSS:2020jck}. The catalog covers nearly the entire sky isotropically, except for the Milky Way plane contaminated by dust and stars. The galaxy redshifts are corrected for peculiar velocities based on the Bayesian Origin Reconstruction from Galaxies (BORG) formalism \citet{Jasche:2012kq,Mukherjee:2019qmm}, up to a redshift of $z = 0.05$. Beyond this redshift range, the impact of peculiar velocity corrections becomes negligible.
In our analysis, we use the pre-computed LOS redshift priors constructed with K-band GLADE+ data, for the distribution of K-band galaxy absolute magnitude is well described by a Schechter luminosity function with parameters $M_{*,K} = -23.39$ and $\alpha_K = -1.09$ adopted from Ref. \cite{Kochanek:2000im}. The magnitude bright and dim cutoffs are set to be $M_{\rm min} = -27.0$ and $M_{\rm max} = -19.0$ respectively, following the GWTC-4 cosmology paper. The LOS priors are pixelized using \texttt{healpix} \cite{Gorski:2004by,Zonca:2019vzt} with \texttt{Nside}=128, so that the pixel size is much smaller than GW localization areas.

We perform our analysis using the GPU-accelerated implementation of \texttt{gwcosmo}\footnote{The background cosmology implemented in \texttt{gwcosmo} assumes the standard $\Lambda$CDM model, and possible modified gravity effects on the Friedmann equation are not taken into account.} \cite{Papadopoulos:2026puy}, which parallelizes the iteration over posterior samples for all GW events through vectorized inference operations. It leads to a speed-up of $\sim 10^3$ times over the old version running on central processing units (CPUs). We additionally perform the inference using the \texttt{nessai} sampler, which incorporates normalizing flows within a nested sampling framework \cite{Williams:2021qyt,Williams:2023ppp}. 

We consider two different priors on $H_0$ in the analysis. One is the narrow prior ${\rm U}(65,77)$, where ${\rm U}$ denotes the uniform prior. This is motivated by the Hubble tension between the CMB measurements \cite{Planck:2018vyg} and the distance ladder measurements \cite{Riess:2019cxk}, and our prior range encompasses both results. The second is a wider prior ${\rm U}(10,120)$, which does not incorporate any prior information from other electromagnetic measurements. In addition, we consider the prior on the spacetime dimension number $D$ to be ${\rm U}(3,12)$, and we choose to sample over $\log(R_c/{\rm Mpc})$ using different prior ranges with a fixed lower bound of $\log(R_c/{\rm Mpc})=2.7$, corresponding to $R_c\simeq  500~{\rm Mpc}$. The adopted prior ranges are motivated by the fact that the LOS redshift prior used in our analysis is truncated at a maximum redshift of $z_{\rm max}=10$. For extreme combinations of cosmological and modified-gravity parameters, such as large $H_0$, small $D$, and small $R_c$, the source redshifts inferred from the GW injection distances can exceed this limit. We therefore restrict the prior ranges to ensure that all injected events remain within the support of the redshift prior. Furthermore, setting the lower bound to $R_c \simeq 500,{\rm Mpc}$ guarantees that departures from GR can be effectively probed, since the majority of GW events are located at distances beyond this scale, where modified-gravity effects become appreciable.

\section{Results}
\label{sec:results}

\begin{figure}
    \centering
    \includegraphics[width=\linewidth]{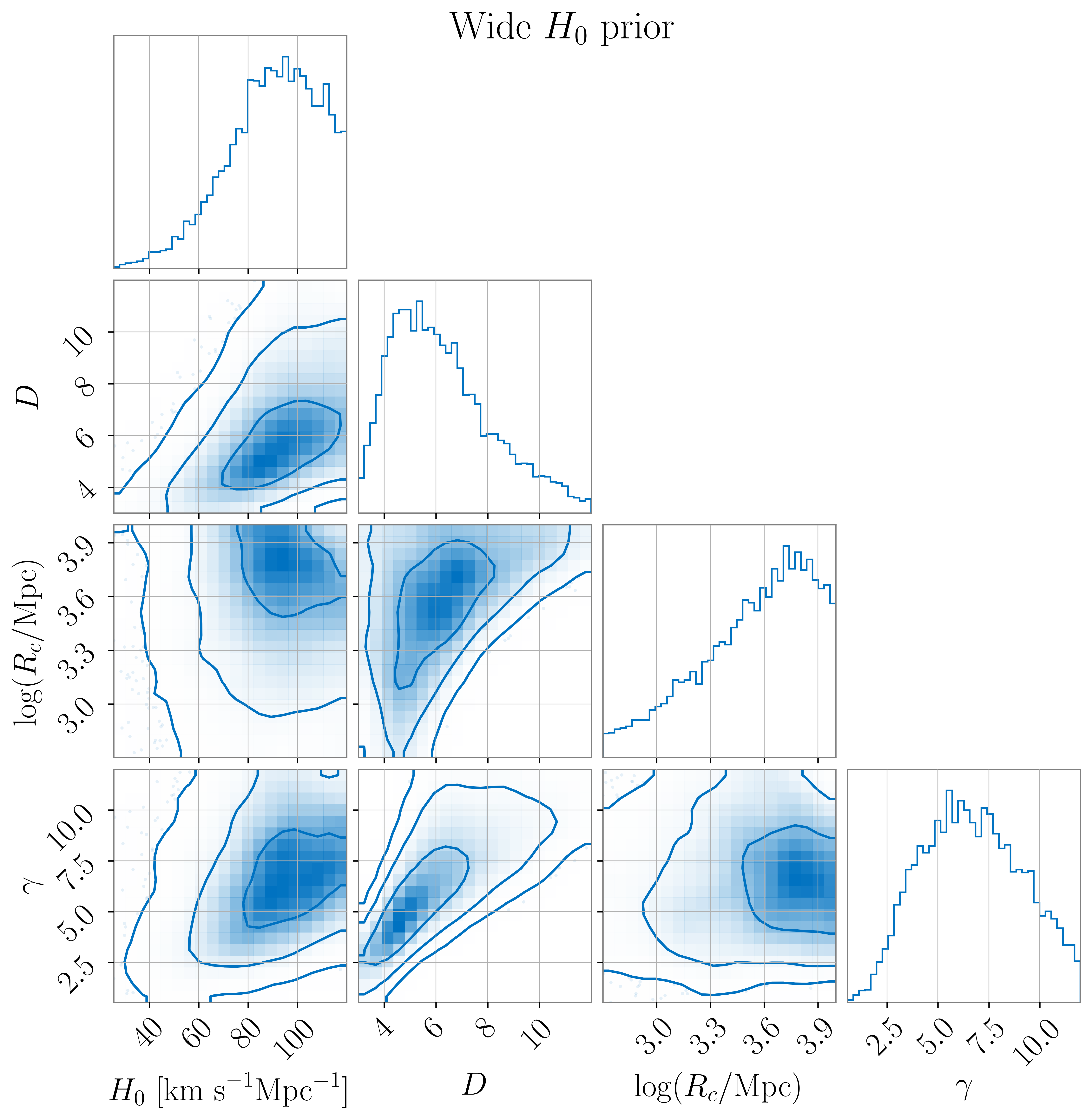}
    \caption{Selected contours for joint posterior probability distribution on $H_0$, $D$, $R_c$ and $\gamma$. The prior on $H_0$ is a wide prior of ${\rm U}(10,120)$. The priors on $D$ and $\log(R_c/{\rm Mpc})$ are ${\rm U}(3,12)$ and ${\rm U}(2.7,4.0)$ respectively.}
    \label{fig:D_logRc4_corner_fullpop_wide_prior}
\end{figure}

We present our inference results in this section. First, we show the selected contours for joint posterior probability distribution  of $H_0$, $D$, $\log(R_c/{\rm Mpc})$ and $\gamma$ (spectral index of merger rate redshift evolution) for the wide $H_0$ prior and the $\log(R_c/{\rm Mpc})$ prior of ${\rm U}(2.7,4.0)$ in Fig. \ref{fig:D_logRc4_corner_fullpop_wide_prior}. The corner plot reveals strong degeneracies between the parameter pairs $H_0$--$D$, $D$--$R_c$, and $D$--$\gamma$. These degeneracies can be understood from the roles they play in shaping the observed GW population. While $H_0$ affects $d_L^{\rm GW}(z)$ through the cosmic expansion history, $D$ and $R_c$ modify $d_L^{\rm GW}(z)$ through deviations from GR at a given redshift. As a result, changes in $H_0$ can be partially compensated by changes in the modified-gravity parameters, leading to strong correlations among them. On the other hand, the degeneracy between $D$ and $\gamma$ arises because the intrinsic merger-rate density evolves with redshift as $\psi(z|\Lambda)\propto(1+z)^\gamma$ (see Appendix \ref{app:pop_mod} for details). Consequently, while $\gamma$ alters the merger-rate redshift evolution, $D$ modifies the inferred redshift distribution through its effect on $d_L^{\rm GW}(z)$. Their competing impacts on the redshift dependence of the observed population therefore give rise to a strong degeneracy. 

In addition, the posterior distributions of $H_0$ and $\log(R_c/{\rm Mpc})$ exhibit significant support near the upper edges of their prior ranges, indicating that the posteriors are prior-limited at the upper bounds. The marginalized posterior distribution of $H_0$ peaks at approximately $90~{\rm km~s^{-1}~Mpc^{-1}}$, driven by its degeneracy with $D$. The marginalized posterior distribution of $D$ gives $D=5.01^{+2.38}_{-1.27}$. Nevertheless, the GR limit of $H_0 \simeq 70~{\rm km,s^{-1},Mpc^{-1}}$ and $D=4$ remain well within the $1\sigma$ credible region of the contour. 

\begin{figure}
    \centering
    \includegraphics[width=\linewidth]{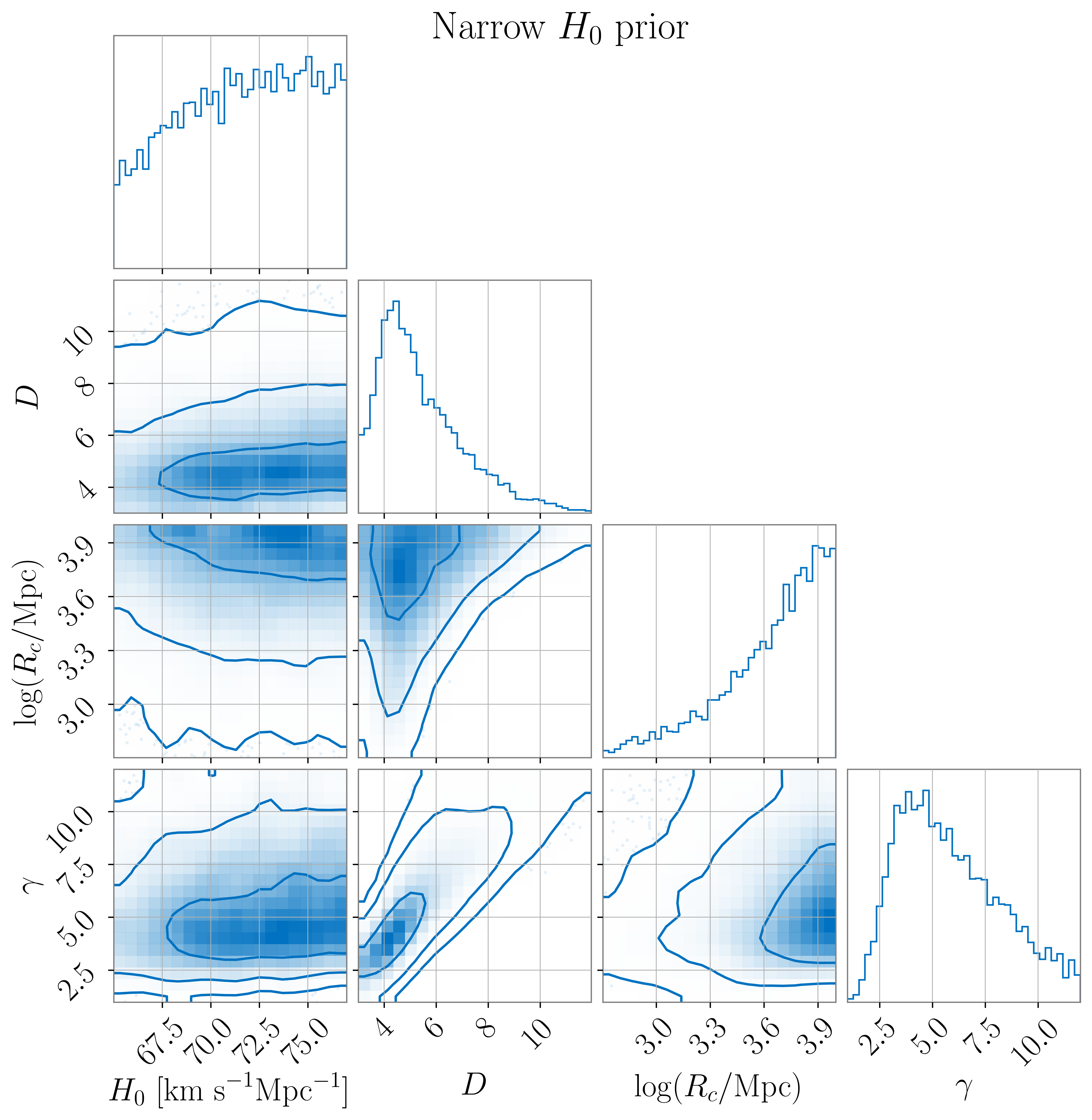}
    \caption{Selected contours for joint posterior probability distribution on $H_0$, $D$, $R_c$ and $\gamma$. The prior on $H_0$ is a narrow prior of ${\rm U}(65,77)$. The priors on $D$ and $\log(R_c/{\rm Mpc})$ are ${\rm U}(3,12)$ and ${\rm U}(2.7,4.0)$ respectively.}
    \label{fig:D_logRc4_corner_fullpop}
\end{figure}

Next, we show in Fig. \ref{fig:D_logRc4_corner_fullpop} the same selected contours for joint posterior probability distribution as in Fig. \ref{fig:D_logRc4_corner_fullpop_wide_prior} except switching to the narrow $H_0$ prior. Since the $H_0$ prior becomes much narrower, the degeneracy between $H_0$ and $D$ is broken, so that the constraint on $D$ becomes significantly stronger, yielding $D=4.38^{+1.91}_{-1.01}$. Although the $D$--$\gamma$ degeneracy persists, the constraint on $\gamma$ is also slightly tightened. However, the posterior samples of $\log(R_c/{\rm Mpc})$ continue to accumulate near the upper prior boundary.

\begin{figure}[t]
    \centering
    \includegraphics[width=\linewidth]{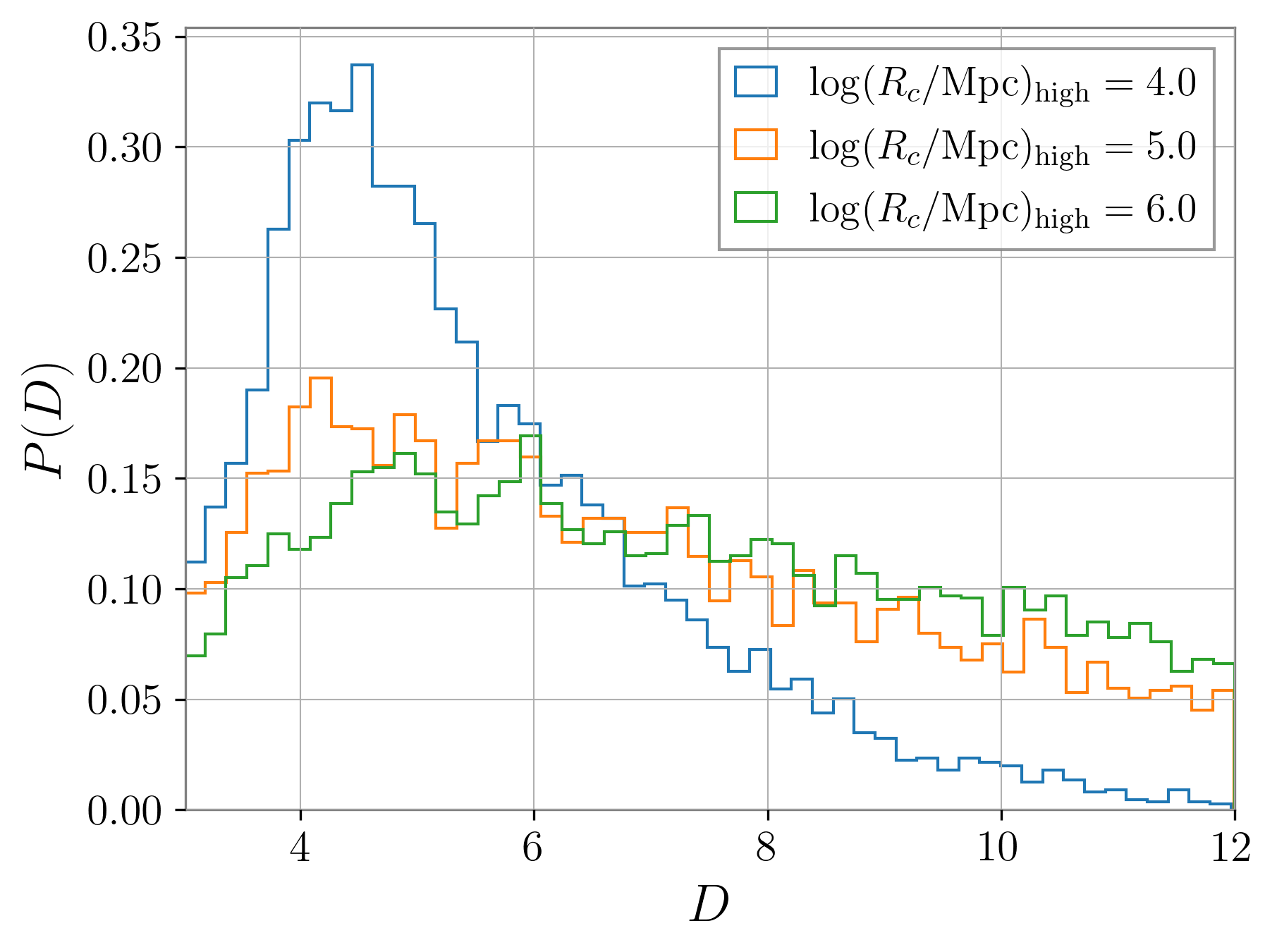}
    \caption{Posterior probability distribution on $D$ with a narrow $H_0$ prior and with different higher bounds for the prior on $\log(R_c/{\rm Mpc})$.}
    \label{fig:D_compare_Rc}
\end{figure}

To investigate the impact of the higher prior bound on $\log(R_c/{\rm Mpc})$, denoted by $\log(R_c/{\rm Mpc})_{\rm high}$, we repeat the analysis using extended uniform priors with $\log(R_c/{\rm Mpc})_{\rm high}=5.0$ and $6.0$, while keeping the priors on all other parameters unchanged and adopting the narrow $H_0$ prior. Fig.~\ref{fig:D_compare_Rc} compares the marginalized posterior distributions of $D$ obtained for the different choices of $\log(R_c/{\rm Mpc})_{\rm high}$. We find that the constraints on $D$ become substantially weaker when the upper bound is increased from $4.0$ to $5.0$ or $6.0$.
To further illustrate this behavior, we present the full joint posterior contours for the different prior choices in Appendix~\ref{app:Rc_full_corner}. In all cases, the posterior distribution of $\log(R_c/{\rm Mpc})$ continues to accumulate near the upper prior boundary, indicating that the parameter remains unconstrained from above. This trend can be understood physically: for larger values of $R_c$, the modifications to GW propagation induced by $D>4$ become increasingly suppressed at the low distances where most GW events are observed, causing the signal to approach the GR limit and reducing the constraining power on both $R_c$ and $D$. On the other hand, as shown in Fig. \ref{fig:D_logRc5_corner_fullpop} and \ref{fig:D_logRc6_corner_fullpop}, since the constraint on $D$ becomes significantly weaker with higher prior bound on $R_c$, the $D$--$\gamma$ degeneracy is alleviated, allowing $\gamma$ to be constrained more tightly.

These results demonstrate that adopting a physically motivated upper prior bound on $\log(R_c/{\rm Mpc})$ is essential for obtaining meaningful constraints on higher-dimensional theories of gravity. The choice of $\log(R_c/{\rm Mpc})_{\rm high}=4.0$ places the crossover scale within a range that is comparable to the distances of the observed GW events, allowing deviations from GR associated with $D>4$ to be effectively probed and thereby yielding informative constraints on $D$. In contrast, when $\log(R_c/{\rm Mpc})_{\rm high}$ is set to values substantially larger than the typical GW event distances, the modified-gravity effects become strongly suppressed across the observed population. As a result, the current GW dataset provides only weak constraints on both $D$ and $R_c$, making such broad prior ranges poorly suited for constraining higher-dimensional theories with present observations.

Conversely, a smaller value of $R_c$ enhances the sensitivity of GW observations to modifications of gravity, leading to significantly tighter constraints on $D$. This effect was previously demonstrated in the analysis of the multimessenger event GW170817 \cite{Pardo:2018ipy}, which obtained $D=3.98^{+0.07}_{-0.09}$ under the assumptions of a fixed Planck-measured value of $H_0$ and $R_c=1~{\rm Mpc}$. Although both their result and ours are consistent with GR, the uncertainty on $D$ in their analysis is approximately 18 times smaller than that obtained here. Similarly, the spectral-siren analysis of the GWTC-3 catalog found $D=3.95^{+0.09}_{-0.07}$ when adopting the Planck value of $H_0$ and fixing $R_c=1~{\rm Mpc}$  \cite{MaganaHernandez:2021zyc}\footnote{We note that the expression for $d_L^{\rm GW}$ adopted in Refs.~\cite{Pardo:2018ipy,MaganaHernandez:2021zyc} appears to be missing an additional factor of $(1+z)$ multiplying $R_c$. Including this factor is necessary to ensure consistency with the definition of the crossover scale in an expanding cosmological background.}. These results highlight the significant impact of the assumed crossover scale on the inferred constraints on higher-dimensional gravity. 

\begin{figure}[t!]
    \centering
    \begin{subfigure}{\linewidth}
    \includegraphics[width=\textwidth]{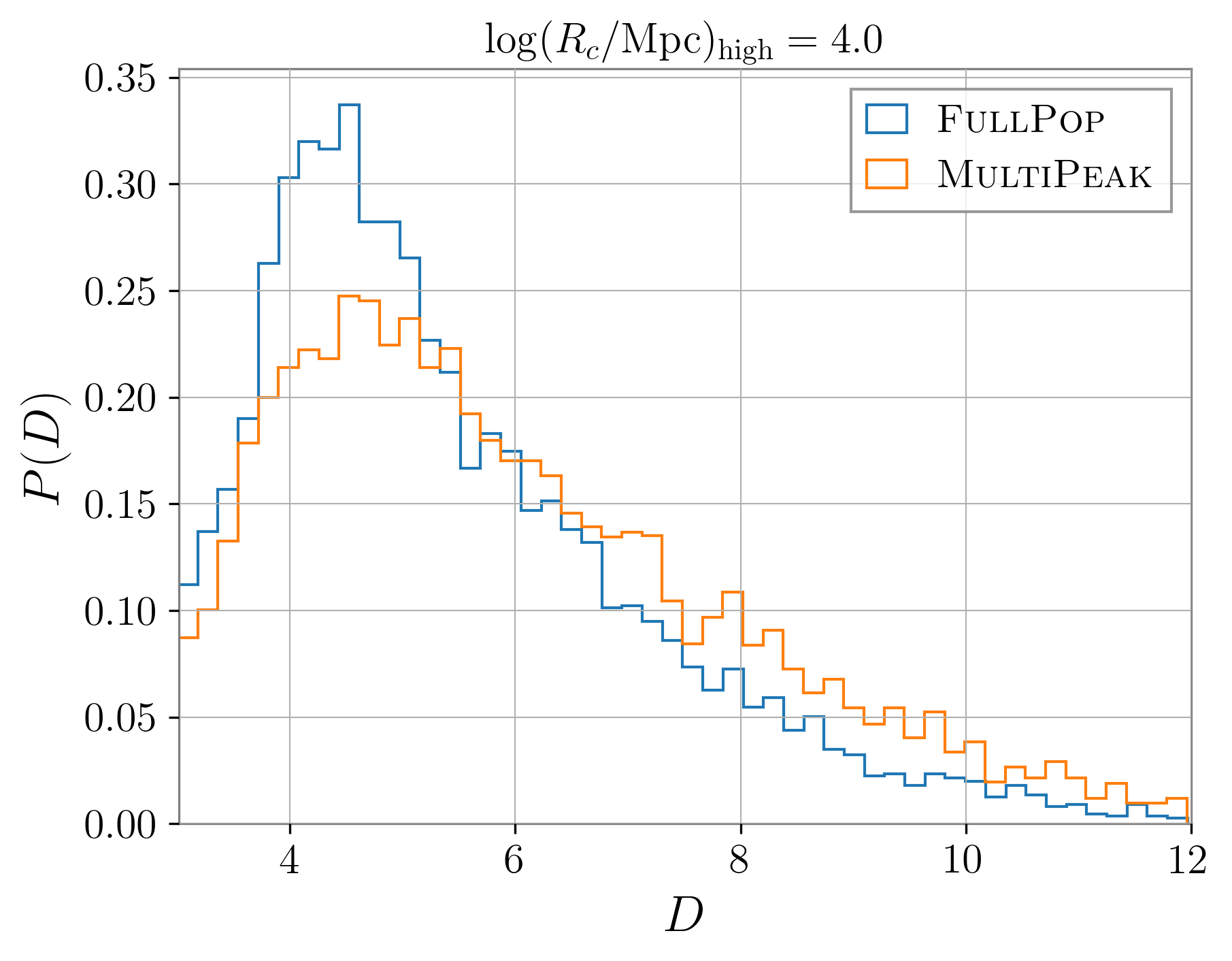}
    \end{subfigure}
    \begin{subfigure}{\linewidth}
    \includegraphics[width=\textwidth]{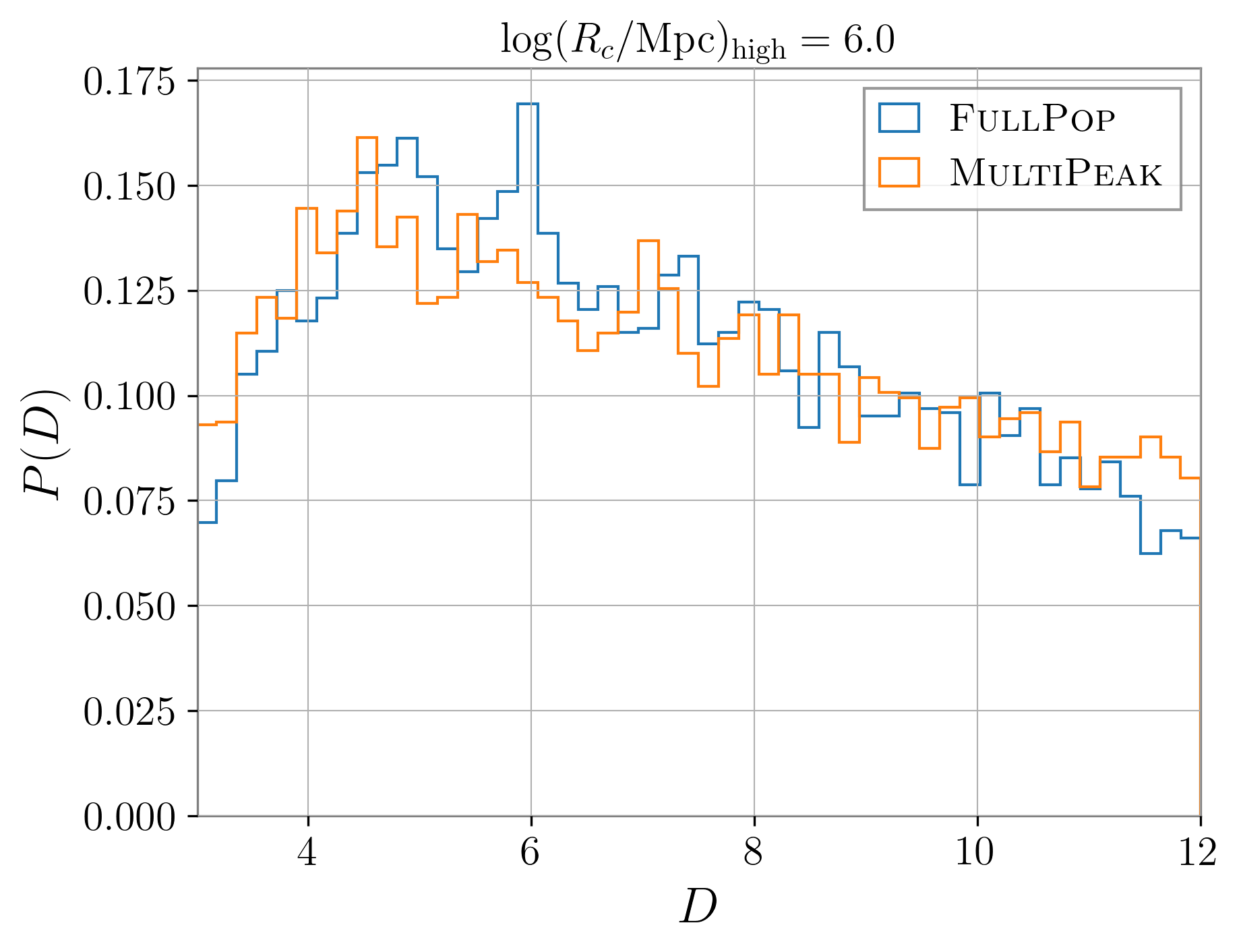}
    \end{subfigure}
    \caption{Comparison of posterior probability distribution on $D$ using 141 CBCs with the \textsc{FullPop}-4.0 model and 137 BBHs with the \textsc{MultiPeak} model, respectively. The upper and the lower panel show the case with $\log(R_c/{\rm Mpc})_{\rm high}=4.0$ and $\log(R_c/{\rm Mpc})_{\rm high}=6.0$, respectively.}
    \label{fig:D_mass_model}
\end{figure}

Joint constraints on $D$ and $R_c$ have previously been obtained using both spectral-siren and dark-siren analyses of the GWTC-3 catalog \cite{MaganaHernandez:2021zyc,Leyde:2022orh,Chen:2023wpj}. Similar to our findings, the posterior distribution of $R_c$ in these studies accumulates near the upper prior boundary, indicating that the crossover scale remains poorly constrained from above. The resulting constraints on $D$ are of a comparable magnitude to those obtained here. However, a direct comparison between our GWTC-4 results and the earlier GWTC-3 analyses is not straightforward, as the underlying mass-population models and selection-effect treatments differ substantially.
In particular, the GWTC-3 analyses adopted the \textsc{PowerLaw+Peak} mass model, whereas the \textsc{FullPop}-4.0 and \textsc{MultiPeak} models used in this work introduce a larger number of hyperparameters. The increased model complexity broadens the posterior distributions and leads to greater uncertainty in the Bayesian inference. Furthermore, the detector sensitivity during the LVK O4 observing run differs from that of O3, resulting in a different detection probability and hence a different selection-effect correction. Consequently, despite the larger number of GW events available in GWTC-4, the constraints on $D$ do not improve significantly relative to those obtained from GWTC-3.

We also compare the constraints obtained using different mass-population models, as shown in Fig. \ref{fig:D_mass_model}. In the upper panel, corresponding to the case with $\log(R_c/{\rm Mpc})_{\rm high}=4.0$, the constraint on $D$ obtained with the \textsc{MultiPeak} model is noticeably weaker than that derived using the \textsc{FullPop}-4.0 model. One possible explanation is that the \textsc{FullPop}-4.0 analysis includes four additional neutron-star-containing events. Because these events are generally located at smaller distances than most BBH mergers, their luminosity distances are measured more precisely, thereby providing greater sensitivity to modified-gravity effects arising from deviations of $D$ from its GR value of 4.
A second possibility is that the neutron-star population introduces an additional feature in the mass distribution that is absent in the BBH-only \textsc{MultiPeak} model. Such a feature may help break degeneracies and improve cosmological constraints, especially given that the constraining power of current GW data is known to depend strongly on the assumed mass model, as we discuss below.
However, the lower panel of Fig. \ref{fig:D_mass_model} shows that the difference between the \textsc{FullPop}-4.0 and \textsc{MultiPeak} constraints becomes much smaller when the upper prior bound is increased to larger values of $\log(R_c/{\rm Mpc})_{\rm high}=6.0$, where the overall constraint on $D$ is weaker. This behavior can be understood from the fact that higher-dimensional effects are strongly suppressed at the relatively short distances of the four neutron-star-containing events. Consequently, including these events has little impact on the combined constraint when the crossover scale is sufficiently large. We therefore conclude that the stronger constraint obtained with the \textsc{FullPop}-4.0 model is more likely driven by the improved distance measurements of these nearby events when $R_c$ is small enough for modified-gravity effects to become appreciable at their distances.

\begin{figure}
    \centering
    \includegraphics[width=\linewidth]{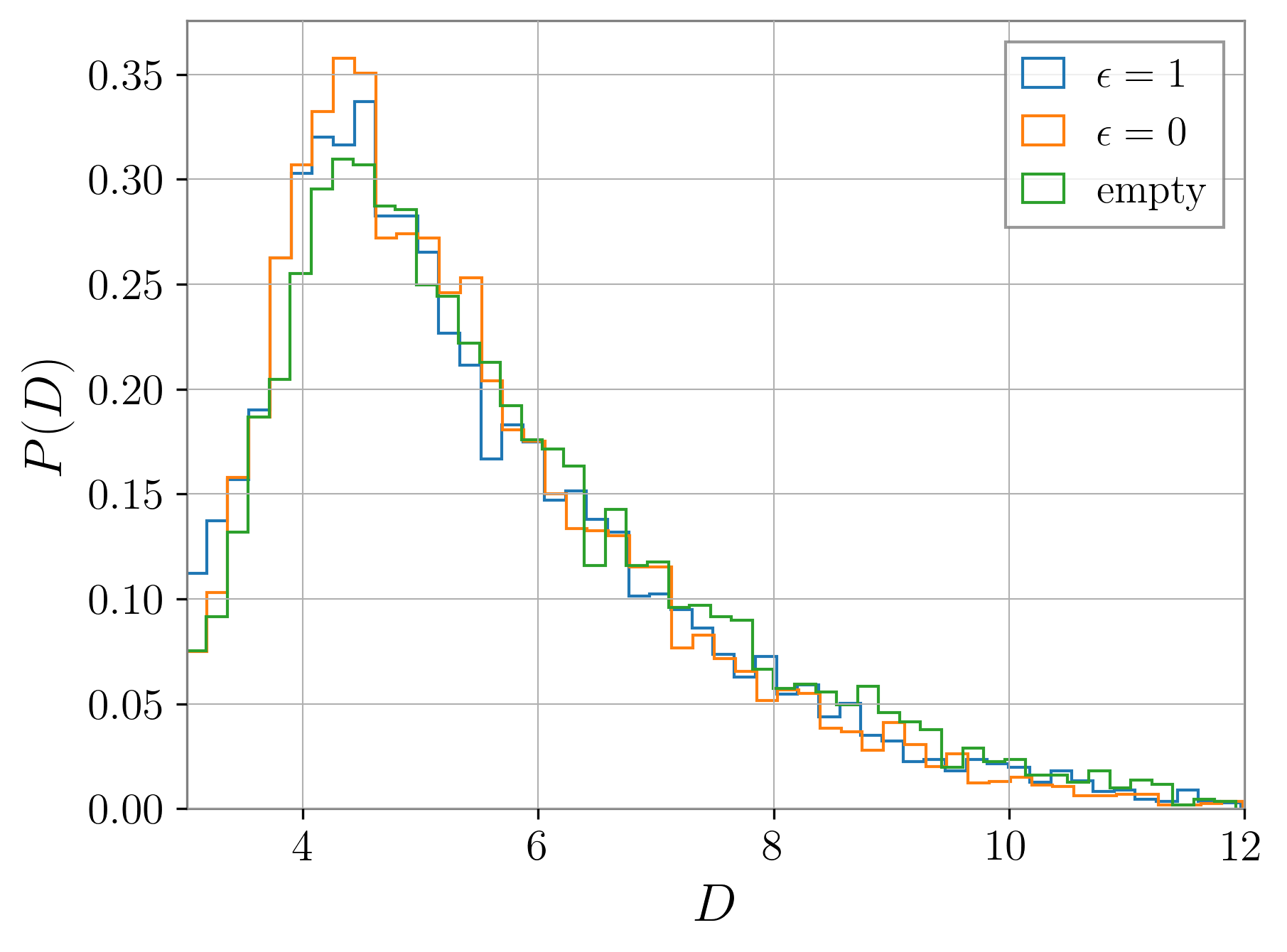}
    \caption{Posterior probability distribution on $D$ with a narrow $H_0$ prior and with different higher bounds for the prior on $\log(R_c/{\rm Mpc})$}
    \label{fig:D_systematic}
\end{figure}

Finally, we investigate the impact of the GLADE+ galaxy catalog on dark-siren constraints of higher-dimensional theories. We compare our default analysis, which employs luminosity weighting, with an alternative analysis using uniform weighting for all galaxies in the catalog while keeping the same hyperparameter priors. Specifically, the two weighting schemes correspond to $\epsilon=1$ and $\epsilon=0$, respectively, in the weighting prescription $w=(L/L_*)^{\epsilon}$. We also consider the limiting case of an empty galaxy catalog. In this scenario, the cosmological information is derived entirely from the population mass model, corresponding to the spectral siren method.
Figure \ref{fig:D_systematic} presents the results for the narrow $H_0$ prior and $\log(R_c/{\rm Mpc})_{\rm high}=4.0$. We find only minor differences among the three cases, indicating that the GLADE+ catalog provides limited additional constraining power for higher-dimensional gravity with the current GW dataset. This can be understood from the fact that the GLADE+ catalog is primarily complete at relatively low distances, extending to approximately $d_L \simeq 130,{\rm Mpc}$, and therefore encompasses only a small fraction of the observed GW events. As a result, its contribution to the overall cosmological inference is limited.
The impact of galaxy-catalog information is expected to increase substantially with deeper surveys that probe higher redshifts, such as the Dark Energy Survey (DES) and the Dark Energy Spectroscopic Instrument (DESI), both of which have already been used in dark-siren measurements of $H_0$ within the $\Lambda$CDM framework \cite{Palmese:2021mjm,DESI:2023fij,Bom:2024afj,McMahon:2026nhi}. Incorporating these catalogs into our analysis may enhance the constraining power on higher-dimensional gravity models. We leave such an investigation to future work.

\section{Conclusions}
\label{sec:conclusions}

In this work, we have presented a dark-siren test of higher-dimensional theories of gravity using the GWTC-4 catalog. We considered a phenomenological modification to the GW luminosity distance motivated by gravity leakage into extra dimensions, characterized by the spacetime dimension number $D$ and the crossover scale $R_c$. By combining 141 CBC events with the GLADE+ galaxy catalog within a hierarchical Bayesian framework, we obtained updated constraints on the spacetime dimension number through GW propagation effects using the dark-siren dataset from the first part of the fourth LVK observation run.

Our analysis reveals that the inferred constraints on the spacetime dimension are strongly correlated with both the Hubble constant and the crossover scale. When adopting a narrow prior on $H_0$, motivated by existing electromagnetic measurements, we obtain a constraint of $D = 4.38^{+1.91}_{-1.01}$ for a prior range $\log(R_c/{\rm Mpc}) \in [2.7,4.0]$. This result is fully consistent with the prediction of General Relativity, $D=4$, and shows no evidence for extra-dimensional gravitational leakage. We also find that the posterior distribution of $R_c$ accumulates near the upper edge of the prior range, indicating that the crossover scale remains poorly constrained by current GW observations.

A key result of this work is the strong dependence of the inferred constraint on the assumed prior range of $R_c$. Increasing the upper prior bound substantially weakens the constraint on $D$, since modified-gravity effects become increasingly suppressed when the crossover scale exceeds the characteristic distances of the observed GW population. This finding is consistent with the results obtained from the GWTC-3 dataset, as well as with forecasts for next-generation GW detectors \cite{Zhu:2026gsu}. This behavior highlights an intrinsic limitation of current datasets: while GW observations are capable of constraining deviations from four-dimensional gravity when the crossover scale is comparable to observed source distances, they provide little information on models with very large crossover scales that closely mimic General Relativity.

We further investigated the impact of population modeling and galaxy-catalog information. The \textsc{FullPop}-4.0 population model yields somewhat tighter constraints than the BBH-only \textsc{MultiPeak} model, primarily due to the inclusion of nearby neutron-star-containing events with more precise distance measurements. In contrast, we find that the choice of galaxy weighting scheme in the GLADE+ catalog has only a minor impact on the inferred constraints. This result is expected because the current catalog is primarily complete at relatively low redshifts and therefore covers only a limited fraction of the observed GW sources.

The present analysis demonstrates both the potential and the current limitations of dark sirens as probes of higher-dimensional gravity. Future improvements are expected from several directions. First, uncertainties are expected to decrease as the number of detected events grows with data from the LVK O4b/c, IR1, and O5 observing runs, provided that the adopted CBC mass model remains adequate to describe the population. Second, detecting high-redshift sources will be important for tightening constraints on modified gravity. The increased detector sensitivity anticipated for O5 and the LIGO A\# upgrade \cite{LIGO:2026gzb} will extend the accessible redshift range, enabling more such detections. Moreover, deeper and more complete galaxy surveys, such as DES, DESI, Euclid, and LSST, are expected to improve the coverage and characterization of potential host galaxies at cosmological distances. These catalogs will be avialable for association with LVK O5 data, which can significantly improve cosmological constraints from dark siren analyses. Furthermore, next-generation GW observatories including the Einstein Telescope and Cosmic Explorer will detect vastly larger source populations with significantly improved localization accuracy and distance measurements. These advances will increase the statistical power of dark sirens and enable much more stringent tests of GW propagation. 

In addition, special classes of events such as golden dark sirens \cite{Nishizawa:2016ood,Borhanian:2020vyr,Gupta:2022fwd,Chen:2024gdn,Chen:2025qsl,Dang:2025vqx,Benetti:2026enj} and strongly lensed gravitational waves \cite{Hannuksela:2020xor,Chen:2024xal,Poon:2024zxn,Chen:2025xeg,Chen:2026htz,Seo:2026eto,Uronen:2026lmu,Chen:2026fsh} may provide particularly powerful probes of modified gravity by reducing host-galaxy ambiguity and supplying additional cosmological information. Furthermore, incorporating additional waveform effects, such as orbital precession, can improve cosmological constraints. Previous studies have shown that strongly precessing binaries can help break the degeneracy between luminosity distance and binary inclination, leading to more precise gravitational-wave distance measurements \cite{Zhu:2023rrx,Zhu:2026gsu}. These improved distance measurements can in turn tighten constraints on cosmological parameters and GW propagation effects.

Overall, our results provide the first GWTC-4 dark-siren constraints on higher-dimensional gravitational-wave propagation and show that current observations remain consistent with four-dimensional General Relativity. As GW detector sensitivity, event statistics, and galaxy surveys continue to improve, dark sirens will become an increasingly powerful tool for testing the fundamental nature of spacetime and gravity on cosmological scales.

\begin{acknowledgments}
We thank Antonio Enea Romano for the internal review of this work. A.C. is supported by the China Postdoctoral Science Foundation under Grant No. 2025M773325, and the National Natural Science Foundation of China (NSFC) under Grant No. E414660101 and 12147103. J.Z. is supported by the NSFC under Grants No.~E414660101 and No.~12147103, and the Fundamental Research Funds for the Central Universities under Grants No.~E4EQ6604X2 and No.~E3ER6601A2. This material is based upon work supported by NSF's LIGO Laboratory which is a major facility fully funded by the National Science Foundation.
We are grateful for computational resources provided by the LIGO Laboratory and supported by the National Science Foundation Grants PHY-0757058 and PHY-0823459.
\end{acknowledgments}

\appendix

\section{Crossover scale in the DGP theory}
\label{app:DGP}

The action of the DGP model is expressed as \cite{Dvali:2000hr}
\begin{align}
    S = & M_5^3 \int_{\cal M} {\rm d}^5x\sqrt{-\gamma}R 
    + \int_{\partial {\cal M}}{\rm d}^4x\sqrt{-g} \nonumber\\
    & \times \left[ -2M_5^3K + \frac{M_4^2}{2}R - \sigma +{\cal L}_{\rm matter} \right],
    \label{eq:high_dimension_action}
\end{align}
where ${\cal M}$ and $\partial{\cal M}$ denote the five-dimensional bulk and the four-dimensional brane, respectively. Here, $\gamma$ represents the determinant of the bulk metric $\gamma_{\mu\nu}$, while $M_4$ and $M_5$ correspond to the four- and five-dimensional Planck scales. The quantity $K=g^{\mu\nu}K_{\mu\nu}$ is the trace of the extrinsic curvature tensor $K_{\mu\nu}$, and $\sigma$ denotes the brane tension. The crossover scale separating four- and five-dimensional gravitational behavior is defined as
\begin{equation}
    R_c \sim \frac{M_4^2}{2M_5^3}.
\end{equation}
Since the hierarchy $M_4 \gg M_5$ is assumed, the crossover scale is typically very large, of the order of the Hubble radius $c/H_0$ \cite{Brown:2005ug}.

Applying the field equations derived from equation \eqref{eq:high_dimension_action} to a FLRW background yields the modified Friedmann equation in the DGP framework 
\begin{equation}
    H^2+\frac{k}{a^2} 
    = \left( \sqrt{\frac{\rho}{3M_P^2}+\frac{1}{R_c^2}}
    +\epsilon\frac{1}{2R_c^2} \right)^2,
    \label{eq:DGP_friedmann}
\end{equation}
where the parameter $\epsilon$ specifies the branch of the DGP solution. In the regime $\rho/M_P^2 \gg 1/R_c^2$, equation \eqref{eq:DGP_friedmann} reduces to the standard Friedmann equation of general relativity. For the branch $\epsilon=-1$, the late-time evolution leads to $\rho \ll M_P^2/R_c^2$, causing the scale factor to diverge and the universe to effectively approach a five-dimensional spacetime. This branch will not be considered further in this thesis. 

In contrast, the $\epsilon=1$ branch describes a transition from conventional four-dimensional FLRW cosmology to a self-accelerating brane phase as the energy density falls below the critical scale $M_P^2/R_c^2$. In this scenario, cosmic acceleration arises without introducing a cosmological constant $\Lambda$ into the gravitational field equations. This feature has motivated significant interest in observational tests of the DGP model, including GW test considered in this work, despite the fact that the self-accelerating branch is known to suffer from ghost instabilities.

Introducing the effective density parameter
\begin{equation}
    \Omega_{R_c} \equiv \frac{1}{4R_c^2H_0^2},
\end{equation}
the Friedmann equation for a spatially flat DGP universe can be rewritten as
\begin{equation}
    H^2(z) = H_0^2
    \left[
        \sqrt{\Omega_{R_c}}
        +\sqrt{\Omega_{R_c}+\Omega_{m,0}(1+z)^3}
    \right]^2.
\end{equation}

In the DGP model, gravity can propagate into the extra-dimensional bulk on scales larger than $R_c$. Consequently, the amplitude of gravitational waves decays more rapidly than predicted by GR, such that the GW luminosity distance can be described by equation \eqref{eq:dL_GW}.

\section{Population model}
\label{app:pop_mod}

\begin{table*}[t]
\centering
\caption*{\textsc{MultiPeak}}
\renewcommand{\arraystretch}{1.25}
\begin{tabular}{ccc}
\hline
Parameter & Description & Prior \\
\hline\hline
$\alpha$ & Spectral index of primary-mass power law & $\mathrm{U}(1.5,12)$ \\
$\beta$ & Spectral index of secondary-mass power law & $\mathrm{U}(-4,12)$ \\
$m_{\min}$ & Minimum primary mass [$M_\odot$] & $\mathrm{U}(2,10)$ \\
$m_{\max}$ & Maximum primary mass [$M_\odot$] & $\mathrm{U}(50,200)$ \\
$\delta_m$ & Smoothing parameter [$M_\odot$] & $\mathrm{U}(10^{-3},10)$ \\
$\mu_g^{\rm low}$ & Location of the first peak [$M_\odot$] & $\mathrm{U}(5,100)$ \\
$\sigma_g^{\rm low}$ & Width of the first peak [$M_\odot$] & $\mathrm{U}(0.4,5)$ \\
$\mu_g^{\rm high}$ & Location of the second peak [$M_\odot$] & $\mathrm{U}(5,100)$ \\
$\sigma_g^{\rm high}$ & Width of the second peak [$M_\odot$] & $\mathrm{U}(0.4,10)$ \\
$\lambda_g$ & Fraction of sources in the peaks & $\mathrm{U}(0,1)$ \\
$\lambda_g^{\rm low}$ & Fraction of sources in the first peak & $\mathrm{U}(0,1)$ \\
\hline
\end{tabular}
\caption{Hyperparameters for the \textsc{MultiPeak} mass model and their priors used for analysis. U (LU) stands for uniform (log-uniform) prior.}
\label{tab:mltp_prior}
\end{table*}

\begin{table*}[t]
\centering
\caption*{\textsc{FullPop}-4.0}
\renewcommand{\arraystretch}{1.25}
\begin{tabular}{ccc}
\hline
Parameter & Description & Prior \\
\hline\hline
$\alpha_1$ & Spectral index of the power law before $b$ & $\mathrm{U}(-4,12)$ \\
$\alpha_2$ & Spectral index of the power law after $b$ & $\mathrm{U}(-4,12)$ \\
$\beta_1$ & Spectral index of the pairing function before $m_{\rm break}$ & $\mathrm{U}(-4,12)$ \\
$\beta_2$ & Spectral index of the pairing function after $m_{\rm break}$ & $\mathrm{U}(-4,12)$ \\
$m_{\min}$ & Minimum primary and secondary mass [$M_\odot$] & $\mathrm{U}(0.4,1.4)$ \\
$m_{\max}$ & Maximum primary and secondary mass [$M_\odot$] & $\mathrm{U}(50,200)$ \\
$\delta_{\min}$ & 1st smoothing parameter of the low mass [$M_\odot$] & $\mathrm{LU}(10^{-2},1)$ \\
$\delta_{\max}$ & 2nd smoothing parameter of the low mass [$M_\odot$] & $\mathrm{LU}(10^{-3},1)$ \\
$\mu_g^{\rm low}$ & Location of the first peak [$M_\odot$] & $\mathrm{U}(5,150)$ \\
$\sigma_g^{\rm low}$ & Width of the first peak [$M_\odot$] & $\mathrm{U}(0.4,10)$ \\
$\mu_g^{\rm high}$ & Location of the second peak [$M_\odot$] & $\mathrm{U}(5,150)$ \\
$\sigma_g^{\rm high}$ & Width of the second peak [$M_\odot$] & $\mathrm{U}(1,10)$ \\
$\lambda_g$ & Fraction of sources in the peaks & $\mathrm{U}(0,1)$ \\
$\lambda_g^{\rm low}$ & Fraction of sources in the first peak & $\mathrm{U}(0,1)$ \\
$m_d^{\rm low}$ & Left side of the dip [$M_\odot$] & $\mathrm{U}(1.5,3)$ \\
$m_d^{\rm high}$ & Right side of the dip [$M_\odot$] & $\mathrm{U}(5,9)$ \\
$\delta_d^{\min}$ & Smoothing of the left side of the dip [$M_\odot$] & $\mathrm{LU}(0.01,2)$ \\
$\delta_d^{\max}$ & Smoothing of the right side of the dip [$M_\odot$] & $\mathrm{LU}(0.01,2)$ \\
$A$ & Amplitude of the dip & $\mathrm{U}(0,1)$ \\
\hline
\end{tabular}
\caption{Hyperparameters for the \textsc{FullPop}-4.0 mass model and their priors used for analysis. U (LU) stands for uniform (log-uniform) prior.}
\label{tab:fullpop_prior}
\end{table*}

\begin{table*}[t]
\centering
\caption*{Madau-Dickinson Merger Rate}
\renewcommand{\arraystretch}{1.25}
\begin{tabular}{ccc}
\hline
Parameter & Description & Prior \\
\hline\hline
$\gamma$ & Slope of the power law before the point $z_p$ & $\mathrm{U}(0,12)$ \\
$\kappa$ & Slope of the power law after the point $z_p$ & $\mathrm{U}(0,6)$ \\
$z_p$ & Redshift turning point between the power laws & $\mathrm{U}(0,4)$ \\
\hline
\end{tabular}
\caption{Hyperparameters for the Madau-Dickinson merger rate model and their priors used for analysis. U stands for uniform prior.}
\label{tab:madau_prior}
\end{table*}

In this appendix, we describe the mass model and redshift distribution of GW events adopted in our analysis.

We first introduce the \textsc{MultiPeak} black hole mass model, in which the joint mass prior is factorized as
\begin{equation}
p(m_1,m_2|\Lambda) = p(m_1|\Lambda),p(m_2|m_1,\Lambda).
\end{equation}
The primary-mass distribution is modeled as
\begin{align}
p(m_1|\Lambda) &= (1-\lambda_g),{\cal B}(m_1|M_{\rm min}, M_{\rm max}, \alpha) \nonumber\\
&\quad + \lambda_g \lambda_{\rm g,low},{\cal G}(m_1|\mu_{\rm g,low}, \sigma_{\rm g,low}) \nonumber\\
&\quad + \lambda_g (1-\lambda_{\rm g,low}),{\cal G}(m_1|\mu_{\rm g,high}, \sigma_{\rm g,high}),
\end{align}
where ${\cal B}$ denotes the broken power-law distribution and ${\cal G}$ represents a Gaussian component. The secondary-mass distribution $p(m_2|m_1,\Lambda)$ is assumed to follow a power-law in the mass ratio with slope $\beta$. The parameter definitions and adopted priors are summarized in Table \ref{tab:mltp_prior}.

The \textsc{FullPop}-4.0 model is a generalized CBC population model that extends the \textsc{MultiPeak} framework to describe the full mass distribution of compact binaries, including neutron stars and black holes, over a broad mass range. It combines a low-mass power-law component with a smooth cutoff for neutron stars, a high-mass black hole component consisting of a power law plus two Gaussian peaks, and a dip feature at the transition region to model the observed mass gap between neutron stars and black holes, with the dip parameters treated as population-level quantities. Another key feature is its flexible joint mass modeling: instead of conditioning the secondary mass on the primary as in the \textsc{MultiPeak} model, it models both masses symmetrically via
\begin{equation}
    p(m_1,m_2|\Lambda) \propto p(m_1|\Lambda)p(m_2|\Lambda)f(m_1,m_2|\Lambda),
\end{equation}
where a pairing function $f(m_1,m_2|\Lambda)$ enforcing $m_1 \ge m_2$ introduces additional freedom in secondary mass distribution. The full expression of the \textsc{FullPop}-4.0 model is presented in GWTC-4 LVK Collaboration papers \cite{LIGOScientific:2025jau,LIGOScientific:2025pvj}. Overall, the model uses 19 parameters as listed in Table \ref{tab:fullpop_prior} with the priors used, which enables a unified and flexible description of the full compact binary population, improving sensitivity to structure in the GW mass spectrum.

We also apply the Madau-Dickinson merger rate redshift evolution model inspired from the Madau-Dickinson star formation rate \cite{Madau:2014bja} to our analysis, which is used by the GWTC-4 cosmology paper \cite{LIGOScientific:2025jau} as well. The merger rate in function of redshift is given by
\begin{equation}
    \psi(z|\Lambda) = [1+(1+z_p)^{-\gamma-k}] \frac{(1+z)^\gamma}{1+\left(\frac{1+z}{1+z_p}\right)^{\gamma+k}},
\end{equation}
with the parameter description and priors used listed in Table \ref{tab:madau_prior}. In particular, the parameter $\gamma$ governs the low-redshift evolution of the merger rate, such that $\psi(z|\Lambda) \propto (1+z)^\gamma$ in the regime $z \ll z_p$, causing degeneracy with cosmological hyperparameters such as $H_0$ and $D$.

\section{Joint posterior contours for different priors on $R_c$}
\label{app:Rc_full_corner}

\begin{figure}[t]
    \centering
    \includegraphics[width=\linewidth]{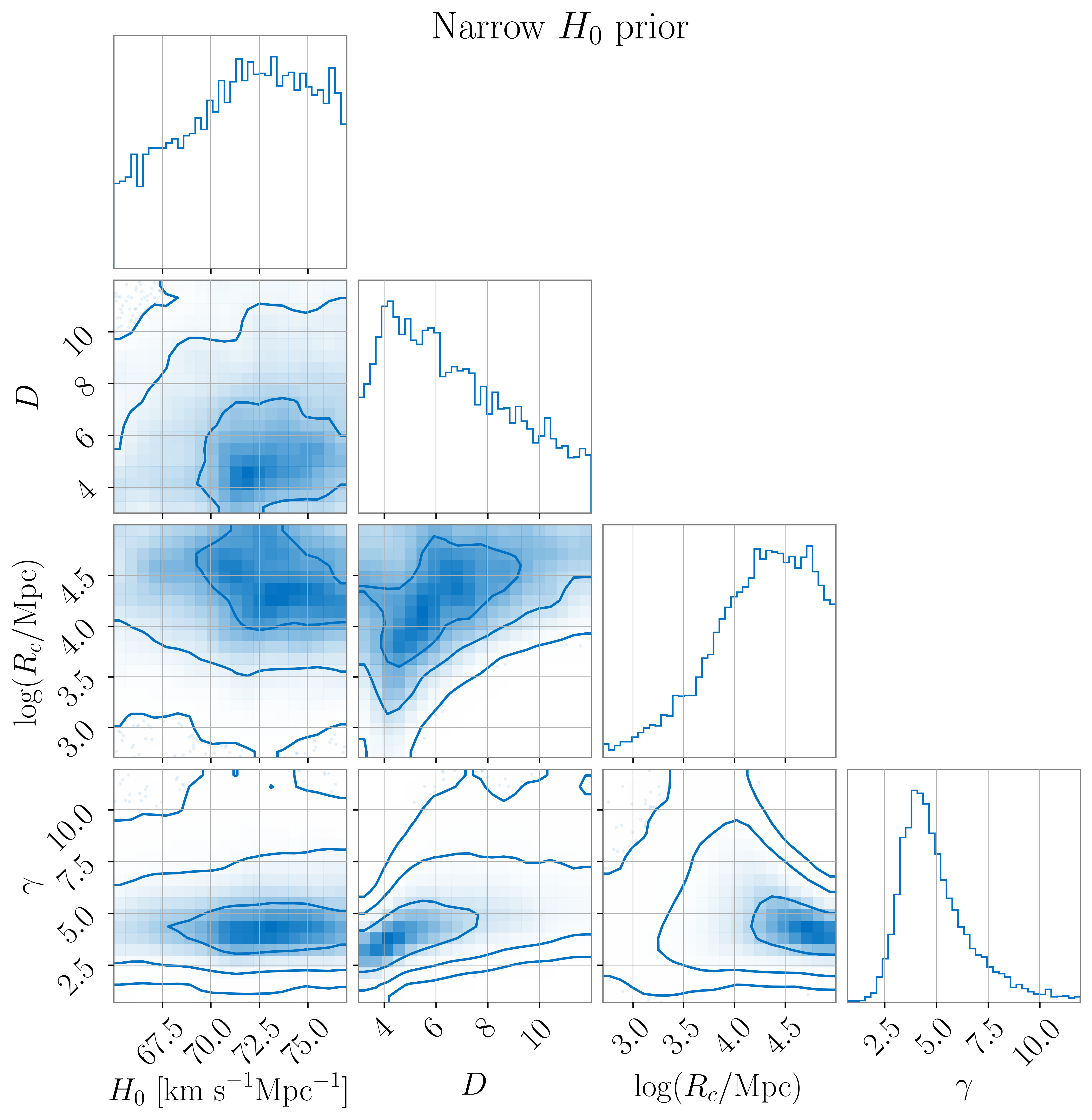}
    \caption{Selected contours for joint posterior probability distribution on $H_0$, $D$, $R_c$ and $\gamma$. The prior on $H_0$ is a narrow prior of ${\rm U}(65,77)$. The priors on $D$ and $\log(R_c/{\rm Mpc})$ are ${\rm U}(3,12)$ and ${\rm U}(2.7,5)$ respectively.}
    \label{fig:D_logRc5_corner_fullpop}
\end{figure}

\begin{figure}[t]
    \centering
    \includegraphics[width=\linewidth]{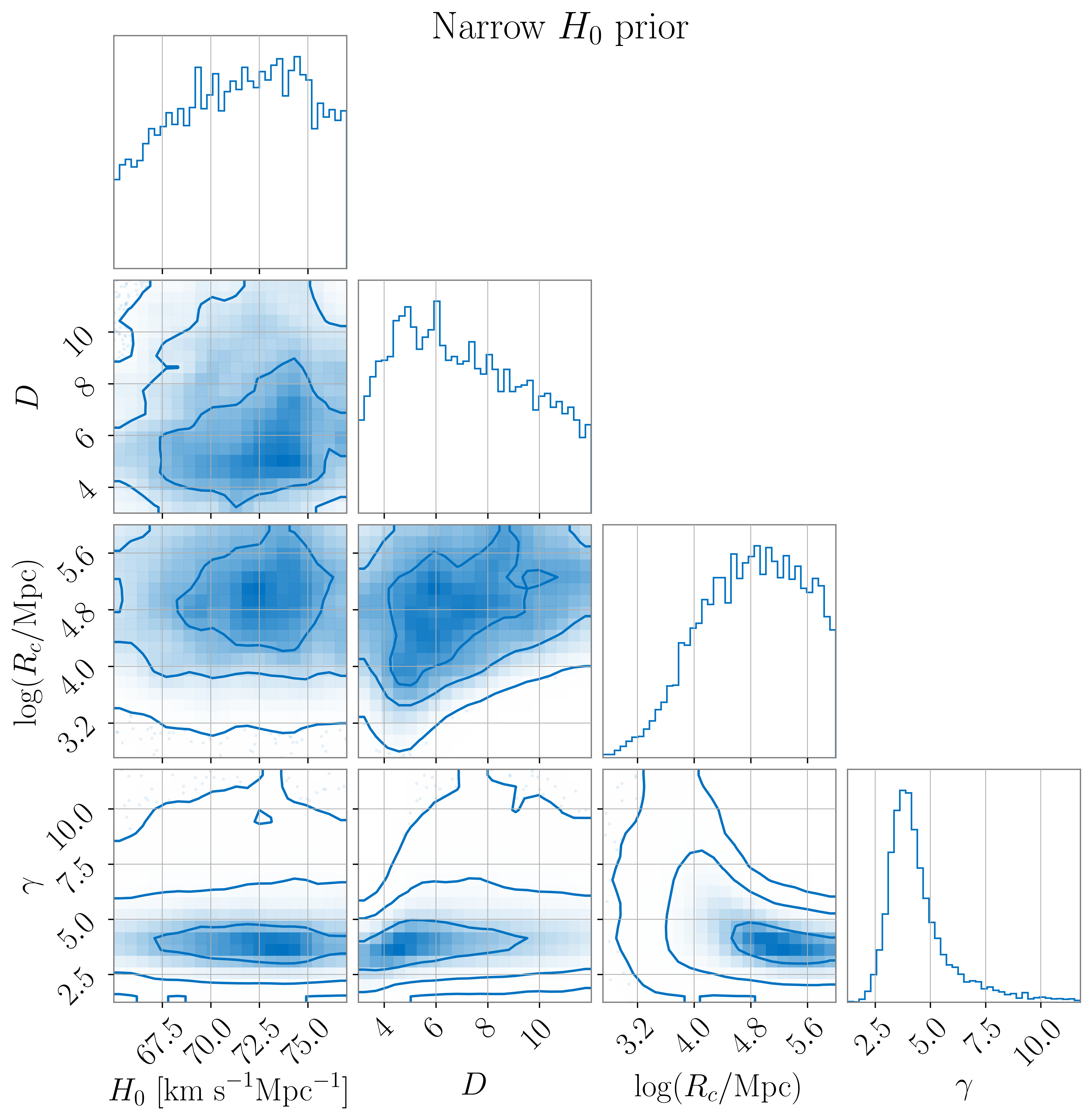}
    \caption{Selected contours for joint posterior probability distribution on $H_0$, $D$, $R_c$ and $\gamma$. The prior on $H_0$ is a narrow prior of ${\rm U}(65,77)$. The priors on $D$ and $\log(R_c/{\rm Mpc})$ are ${\rm U}(3,12)$ and ${\rm U}(2.7,6)$ respectively.}
    \label{fig:D_logRc6_corner_fullpop}
\end{figure}

In this section, we present the selected contours for joint posterior distribution of $H_0$, $D$, $\log(R_c/{\rm Mpc})$ and $\gamma$ for different priors with $\log(R_c/{\rm Mpc})_{\rm high}=5.0$ and $\log(R_c/{\rm Mpc})_{\rm high}=6.0$ in Fig. \ref{fig:D_logRc5_corner_fullpop} and Fig. \ref{fig:D_logRc6_corner_fullpop}, respectively. Both cases use the narrow $H_0$ prior and the \textsc{FullPop}-4.0 mass model, with luminosity weighting for the GLADE+ galaxy catalog. The contours can be directly compared to Fig. \ref{fig:D_logRc4_corner_fullpop}, and the comparison between the marginalized posterior distributions on $D$ in Fig. \ref{fig:D_compare_Rc} are taken from Fig. \ref{fig:D_logRc5_corner_fullpop} and \ref{fig:D_logRc6_corner_fullpop}.

\bibliography{reference}{}
\bibliographystyle{apsrev4-2}

\end{document}